# Lumped and Distributed Parameter SPICE Models of TE Devices Considering Temperature Dependent Material Properties

D. Mitrani, J. Salazar, A. Turó, M. J. García, and J. A. Chávez
Electrical Engineering Department, Universitat Politècnica de Catalunya,
Barcelona, Spain. Email: mitrani@eel.upc.edu

***Abstract*- Based on simplified one-dimensional steady-state analysis of thermoelectric phenomena and on analogies between thermal and electrical domains, we propose both lumped and distributed parameter electrical models for thermoelectric devices. For lumped parameter models, constant values for material properties are extracted from polynomial fit curves evaluated at different module temperatures (hot side, cold side, average, and mean module temperature). For the case of distributed parameter models, material properties are calculated according to the mean temperature at each segment of a sectioned device. A couple of important advantages of the presented models are that temperature dependence of material properties is considered and that they can be easily simulated using an electronic simulation tool such as SPICE. Comparisons are made between SPICE simulations for a single-pellet module using the proposed models and with numerical simulations carried out with Mathematica software. Results illustrate accuracy of the distributed parameter models and show how inappropriate is to assume, in some cases, constant material parameters for an entire thermoelectric element.**

## I. Introduction

Thermoelectric modules (TEM) are solid state devices capable of use either in Peltier mode for transporting heat or in Seebeck mode for electrical power generation [1]-[3]. Despite their low efficiency with respect to traditional devices, TEM's present distinct advantages as far as compactness, precision, simplicity and reliability. Applications of thermoelectric (TE) devices cover a wide spectrum of product areas. These include equipment used in military, aerospace, medical, industrial, consumer, and scientific institutions.

As applications for TE devices increase, both manufacturers and users are facing the problem of developing simple yet accurate models for them. Simulations are usually performed with mathematical software by means of numerical methods [4]-[7] or with electronic and thermal simulators that separately solve the electrical and thermal parts of the model. Alternatively, another methodology is to make use of the analogies between electrical and thermal domains and perform the simulation of the device with an electronic simulation tool such as SPICE. An important benefit of such approach is that both electrical and thermal phenomena can be simulated with a common tool, thereby simplifying simulation of the overall system performance, including control electronics as well as thermal elements.

In this work, we propose both lumped and distributed parameter electrical models for TE devices. A significant novelty of the presented models is that temperature dependence of material parameters is considered. To compare the different models, simulations were performed for a TEM working in Peltier mode for the two more extreme cases, i.e., maximum temperature difference and maximum cooling power. To validate the models, mathematical software was used to obtain a numerical solution to the thermoelectric problem taking into account temperature dependence of material parameters.

## II. TEM Description and Formulae

The basic unit of a TEM is a thermocouple. As illustrated in Fig. 1(a), a thermocouple consists of a p-type semiconductor pellet and an n-type semiconductor pellet joined by metal interconnects. The two pellets of each couple and the many couples in a thermoelectric device are connected electrically in series but thermally in parallel and sandwiched between two ceramic plates, as seen in Fig. 1(b). Nevertheless, if the contribution of the metal interconnects is ignored, there is no loss of generality in analyzing a single couple or a single pellet.

When operated in Peltier mode, in order to pump heat from one side of the TEM to the other by means of an electrical current, four energy conversion processes take place in the pellets: Joule heating, Seebeck power generation, Peltier effect, and Thomson effect. These processes, in conjunction with thermal conduction, determine the performance of the module and are governed by temperature dependent material parameters: Seebeck coefficient, $s$, thermal conductivity, $\kappa$, and electrical resistivity, $\rho$. However, by assuming these parameters constant, one-dimensional (1-D) steady-state analysis leads to the widely used equations for TEM's that we review next.

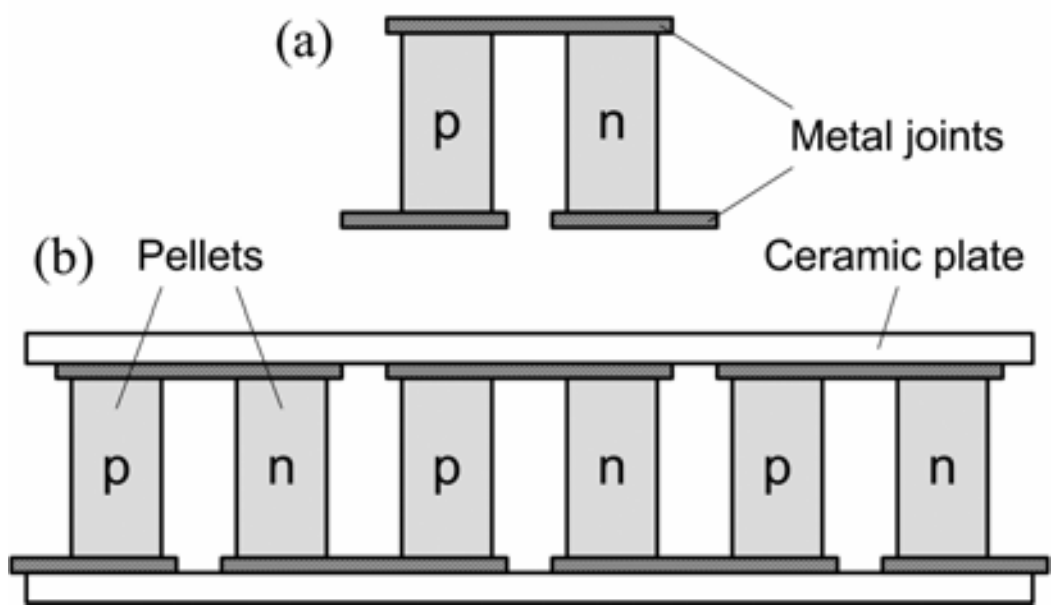

Fig. 1. Schematic of a thermoelectric module, (a) basic unit, and (b) multi-thermocouple module.

Heat absorbed at the cold junction of the module, $Q_c$, and heat released at hot junction, $Q_h$, are respectively given by

$$Q_c = IST_c - \tfrac{1}{2} I^2 R - K\left(T_h - T_c\right), \tag{1}$$

$$Q_h = IST_h + \tfrac{1}{2} I^2 R - K\left(T_h - T_c\right). \tag{2}$$

Where $I$ is the electrical current, $T_c$ and $T_h$ are the module cold and hot side temperatures. For a module made up of $N$ thermocouples with pellets of length $L$ and cross-sectional area $A$, the total Seebeck coefficient, $S$, serial electrical resistance, $R$, and parallel thermal conductance, $K$, are given by

$$K = 2N\frac{\kappa A}{L}, \qquad R = 2N\frac{\rho L}{A}, \qquad S = 2N s\,. \tag{3}$$

Electrical power consumed by the TEM is not simply Joule power. The external current source must also work against Seebeck voltage, and is equal to the difference between heat flow at the hot side and heat flow at the cold side,

$$P_e = Q_h - Q_c = I^2 R + IS\left(T_h - T_c\right). \tag{4}$$

Finally, from the corresponding 1-D expression [8] for temperature distribution along the pellets, $T(x)$, given by

$$T(x) = -\frac{I^2 R}{2KL^2}x^2 + \left(\frac{T_h - T_c}{L} + \frac{I^2 R}{2KL}\right)x + T_c\,, \tag{5}$$

The mean temperature, $T_m$, of the pellets is calculated as

$$T_m = \frac{1}{L}\int_0^L T(x)\,dx = \frac{T_h + T_c}{2} + \frac{1}{12}\frac{I^2 R}{K}. \tag{6}$$

This temperature will serve for some of the models described later, where temperature dependent parameters are calculated according to it.

## III. Steady State Electrical Models

Equations (1)-(4) are widely used as building blocks for a variety of thermoelectric device models, including electrical models [9]-[14]. For these particular types of models, all thermal processes are described in electrical terms using the well-known analogies between electrical and thermal domains described in Table I. According to these analogies,

TABLE I

Analogies between thermal and electrical variables

| **Thermal variable** | **Electrical variable** |
|---|---|
| Heat Flow, $Q$ (W) | Current flow, $I$ (A) |
| Temperature, $T$ (K) | Voltage, $V$ (V) |
| Thermal resistance $R_{th}$ ($W^{-1}K$) | Electrical resistance, $R$ ($\Omega$) |
| Thermal mass, $C_{th}$ ($J{\cdot}K^{-1}$) | Electrical capacity, $C$ (F) |

the thermo-electrical behavior of a TEM can be modeled as an electrical network composed of electrical current sources, voltage sources, resistors, and capacitors. The resulting network can then be simulated by means of electronic circuit simulators such as SPICE.

We shall next describe a steady-state lumped parameter electrical model of a TEM assuming constant material properties, and then proceed to report a distributed parameter electrical model with discrete temperature dependent material properties.

### A. Lumped Parameter Model

Thermoelectric devices can be modeled by a three-port system consisting of two thermal ports and one electrical port, see Fig. 2. Where, referring to the analogies of Table I, thermal ports voltages $T_c$ and $T_h$ correspond to temperature at the cold and hot junctions, while currents $Q_c$ and $Q_h$ represent absorbed and released heat at the cold and hot junctions. At the electrical port, $V$ is the total voltage across the TEM's terminals, and $I$ is the supplied electrical current.

According to expressions for $Q_c$, $Q_h$, and $P_e$, see (1)-(4), Chavez *et a1* [9] have proposed the electrical three-port model of a thermoelectric device shown in Fig. 3 (with $P_e = SIT_c - ½ I^2R$ ). This model clearly illustrates the thermoelectric behavior of a TEM. Within the thermal ports, cooling power and input electrical power are easily readable through current sources $P_x$ and $P_e$, while heat conduction is observed through the corresponding thermal resistance, $R_{th} = K^{-1}$. In the electrical port, overall device voltage is composed of Seebeck voltage, $V_s$, and voltage drop due to module's electrical resistance, $R$.

If boundary conditions of the first kind are applied (temperatures at both ends of the module are known) when simulating the electrical three-port model of Fig. 3, voltage sources have to be connected to thermal ports $T_c$ and $T_h$. If mixed boundary conditions are required (temperature and heat flow at the same or opposite sides are known), one of the voltage sources ($T_c$ or $T_h$) must be changed for the appropriate current source ($Q_c$ or $Q_h$). Electrical power supply is included by connecting either a current or a voltage source to the electrical port.

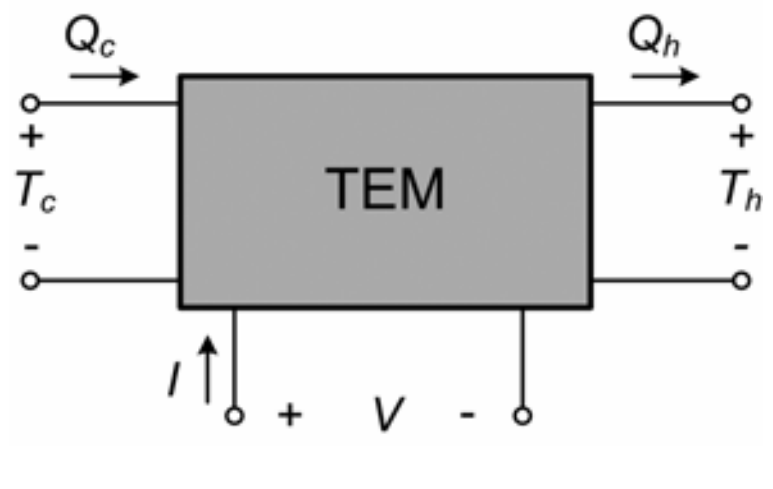

Fig. 2. Three-port block model of a TEM, consisting of two thermal ports and one electrical port.

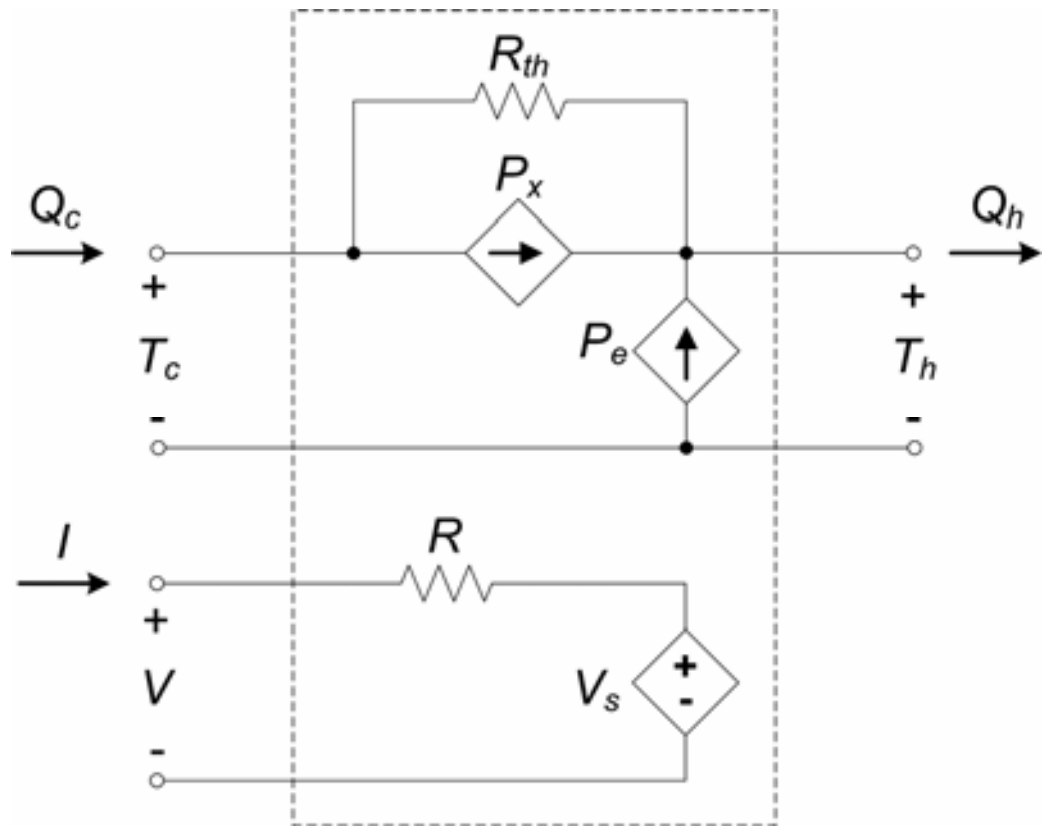

Fig. 3. Steady-state lumped parameter three-port electrical model of a TEM with fixed material properties.

Values used for the constant material properties of the model can be chosen according to different criteria. The simplest method consists in assuming all parameters equal to values at known temperature $T_h$. A more accurate method determines the parameters from mean module temperature, $T_m$. For this method, modifications must be made to the electrical model of Fig. 3. A SPICE voltage-controlled-voltage-source (VCVS) is added in order to determine $T_m$ according to (6). Electrical resistance, $R$, and thermal conductance, $K = R_{th}^{-1}$, are substituted for VCVS's $V_r$ and $T_{con}$ to simulate the corresponding voltage drop by means of expressions that include polynomial approximations for $R(T_m)$ and $K(T_m)$. These approximations, together with the corresponding approximation of Seebeck coefficient, $S(T_m)$, are also used in expressions for $P_e$ and $P_x$. The resulting electrical model is shown in Fig. 4, where VCVS's $S(T_m)$, $R(T_m)$, and $K(T_m)$ are added to monitor the values of these parameters. Similarly, material parameters can also be determined from cold side temperature, $T_c$, or average temperature, $T_{avg} = ½(T_h+T_c)$.

### B. Distributed Parameter Models

Lumped parameter models are only accurate as long as the thermoelectric properties do not vary significantly over the length of the pellets. Hence, if the pellets are divided into many small segments, each segment would be closer to meeting such criteria. Under this condition, material properties for a given segment are assumed to be constant over the small temperature gradient across it. Fig. 5 illustrates a distributed parameter electrical model of a TEM divided for simulation into three segments, where material properties are calculated according to the mean temperature of each segment. To simplify the model, a sub-circuit has been used for each finite element.

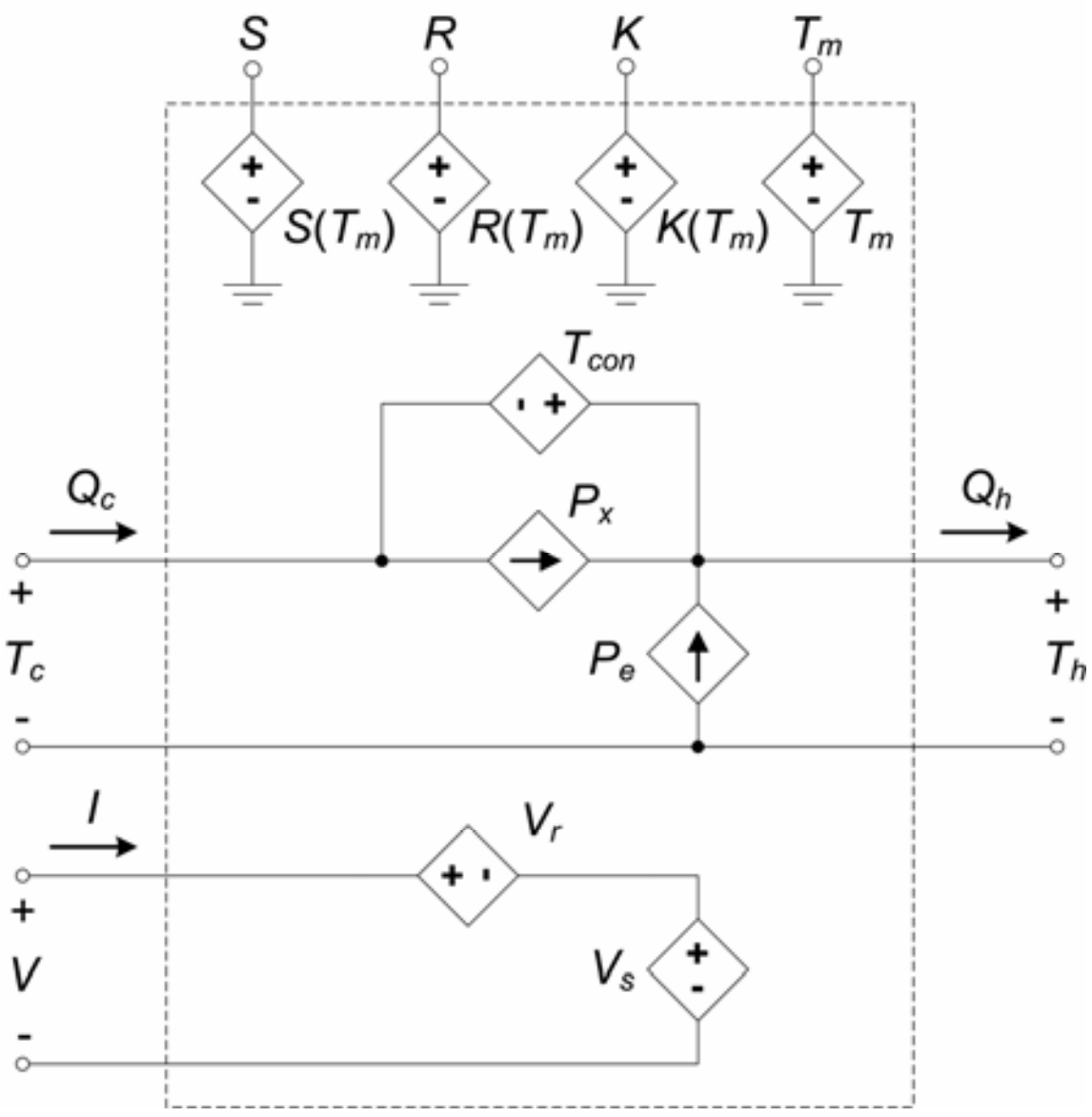

Fig. 4. Steady-state lumped parameter three-port electrical model of a TEM with material properties calculated according to mean temperature, $T_m$.

## IV. Simulation Setup and Results

In this section, comparisons are made between SPICE simulations carried out with the proposed lumped and distributed parameter electrical models for TE devices and with a numerical simulation carried out with Mathematica software [8]. To clarify the results presented hereafter, we shall use the nomenclature presented in Table II to refer to the different TEM models.

### A. Simulation Setup

In order to estimate the value of the temperature dependent parameters according to $T_h$, $T_m$, $T_{avg}$, or $T_c$ for lumped parameter models, or to mean temperature of each segment in a distributed parameter model, as well as for numerical simulation, polynomial approximations of temperature dependent material parameters are required. In this work, we have used the experimentally measured properties of $(Bi_{0.5}Sb_{0.5})_2Te_3$ presented in [15].

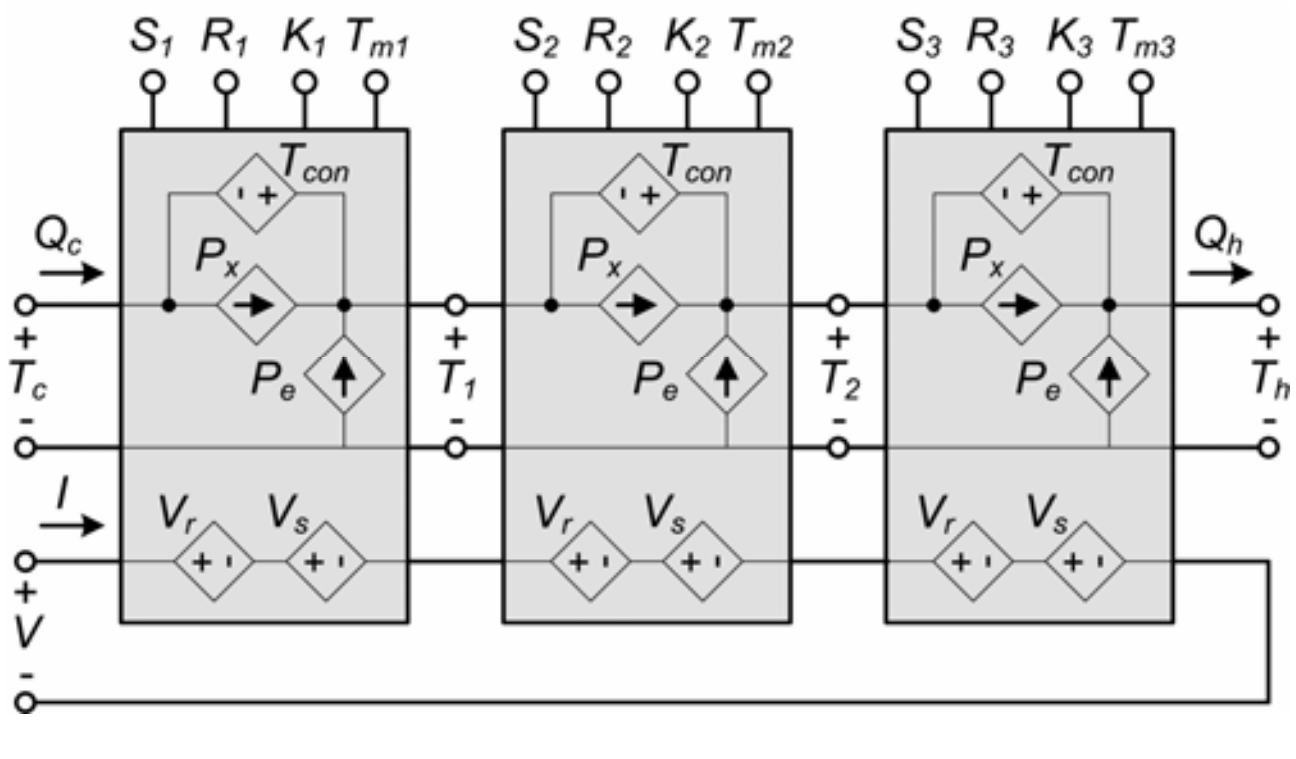

Fig. 5. Steady-state distributed parameter electrical model of a segmented TEM, divided for simulation into three segments represented by the corresponding sub-circuits.

TABLE II
NOMENCLATURE USED TO REFER TO THE DIFFERENT TEM MODELS EMPLOYED THROUGHOUT THIS WORK

| Ref. | Model Description |
|---|---|
| A | Lumped param. SPICE model with param. eval. at $T_h$ |
| B | Lumped param. SPICE model with param. eval. at $T_m$ |
| C | Lumped param. SPICE model with param. eval. at $T_{avg}$ |
| D | Lumped param. SPICE model with param. eval. at $T_c$ |
| E | Distributed param. SPICE model with 3 finite elements |
| F | Dist. param. SPICE model with 10 finite elements |
| G | Dist. param. SPICE model with 20 finite elements |
| H | Numerical Simulation |

For all simulations presented here, device parameters correspond to a single $(Bi_{0.5}Sb_{0.5})_2Te_3$ pellet of cross-sectional area $A$=10 mm$^2$ and length $L$=1 mm. For comparative reasons with [8] and to resemble a practically relevant situation, all simulations where made for a fixed hot side temperature $T_h$=300 K.

### *B. Simulation Results*

As mentioned before, of all the possible operating conditions for a TEM, we will limit our discussion to the particular cases of temperature difference, $\Delta T$, at zero heat absorption ($Q_c$=0 W), and cooling power, $Q_c$, at zero temperature difference ($\Delta T$=0ºC). Comparisons between the different models for these two cases over a broad electrical current range are shown in Figs. 6(a) and 7(a),[1] where the secondary axis shows the relative error with respect to numerical simulation (model H). As expected, at low currents where the temperature profile is relatively flat, there is not much difference between any of the models. However, as electrical current increments and the temperature profile becomes more pronounced, differences between models become noticeable. As can be seen in the expanded graphs of Figs. 6(b) and 7(b), there is an electrical current that produces a maximum temperature difference, $\Delta T_{max}$, and maximum cooling power, $Q_{cmax}$. Values of $\Delta T_{max}$ and $Q_{cmax}$ and corresponding electrical currents $I_{\Delta Tmax}$ and $I_{Qcmax}$ for each model are summarized in Table III. Clearly, the models that best resemble the highly realistic numeric simulation results are the distributed parameter models. Furthermore, according to results for this particular single-pellet TEM, using ten finite elements proves to be sufficient to produce accurate results (0.03% relative error at $I_{\Delta Tmax}$ and 0.01% at $I_{Qcmax}$). With concerns to lumped parameter models, using material parameters evaluated at mean temperature $T_m$ (model B) is the most accurate (3.2% relative error) for the case of zero temperature difference (where $T_c$=$T_h$=$T_{avg}$), and provides very similar results to model C for the case when $Q_c$=0 W (0.2% relative error). For both cases, model D has the largest error.

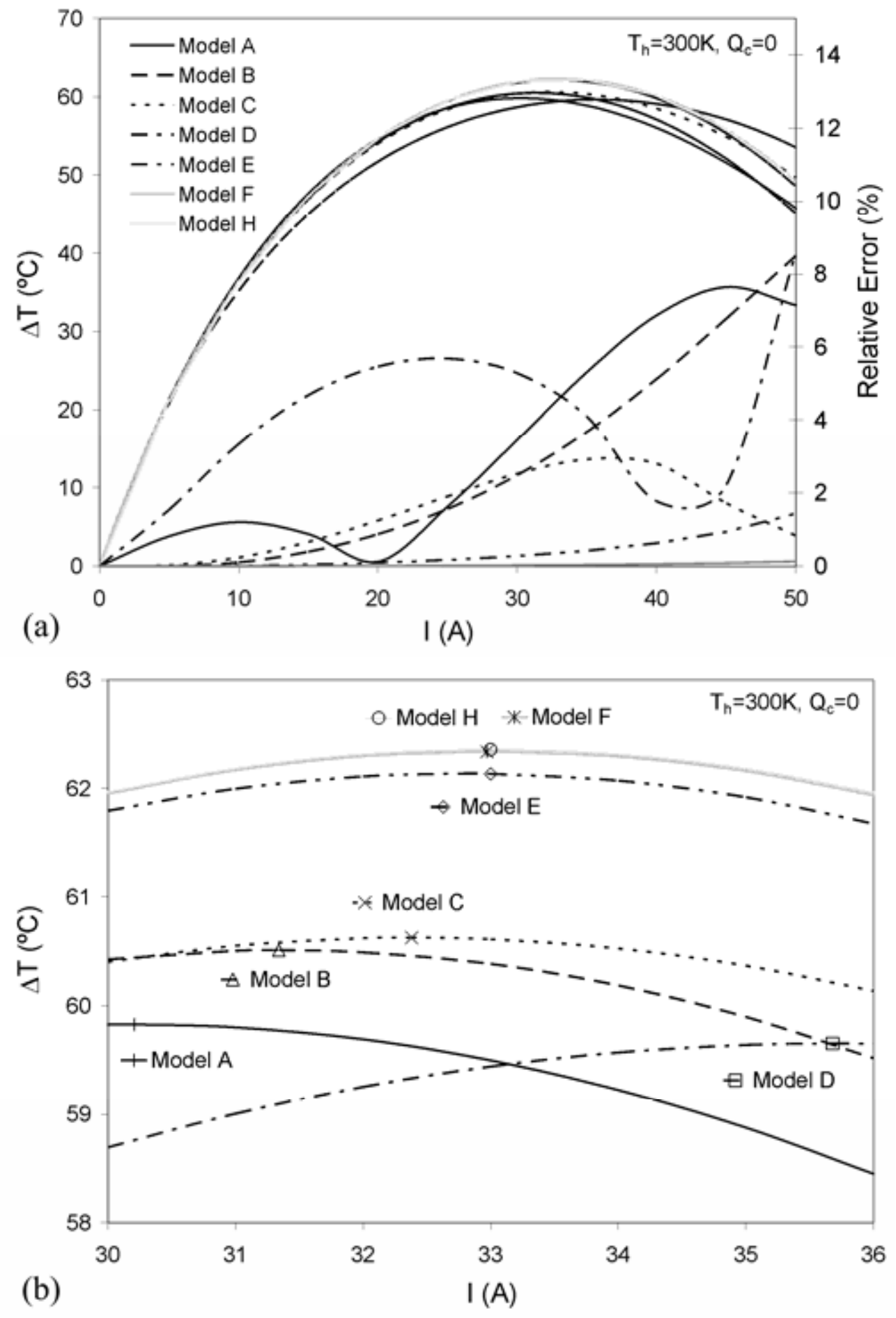

Fig. 6. Prediction comparisons between the different models in Table II for temperature difference vs. electrical current at $Q_c$=0 W, (a) broad current range, including a secondary axis to show the relative error with respect to model H, and (b) expanded data near maximum current.

To better understand the variations presented between the different models, simulations where carried out to determine the corresponding temperature distribution, $T(x)$, for each of the $\Delta T_{max}$ and $Q_{cmax}$ cases presented in Table III (an equivalent distributed parameter model for models A, B, C and D was used applying constant material properties to each finite element).

The resulting temperature profiles are shown in Figs. 8 (in practice, a TEM should operate between these two curves).

TABLE III
COMPARISONS BETWEEN $\Delta T_{max}$ AND $Q_{cmax}$ VALUES

| Model | $\Delta T_{max}$ (ºC) | $I_{\Delta Tmax}$ (A) | $Q_{cmax}$ (W) | $I_{\Delta Tmax}$ (A) |
|---|---|---|---|---|
| A | 59.83 | 30.21 | 1.237 | 37.74 |
| B | 60.51 | 31.34 | 1.230 | 36.95 |
| C | 60.63 | 32.38 | 1.237 | 37.74 |
| D | 59.65 | 35.68 | 1.237 | 37.74 |
| E | 62.14 | 32.77 | 1.231 | 37.04 |
| F | 62.34 | 32.97 | 1.232 | 37.14 |
| G | 62.35 | 32.99 | 1.232 | 37.15 |
| H | 62.36 | 32.99 | 1.232 | 37.15 |

[1] Do to the great similarity in predictions by all distributed parameter models and numerical simulation, models E and F where excluded from Figs. 6-9 to add clarity. However results for these models are included in Table III.

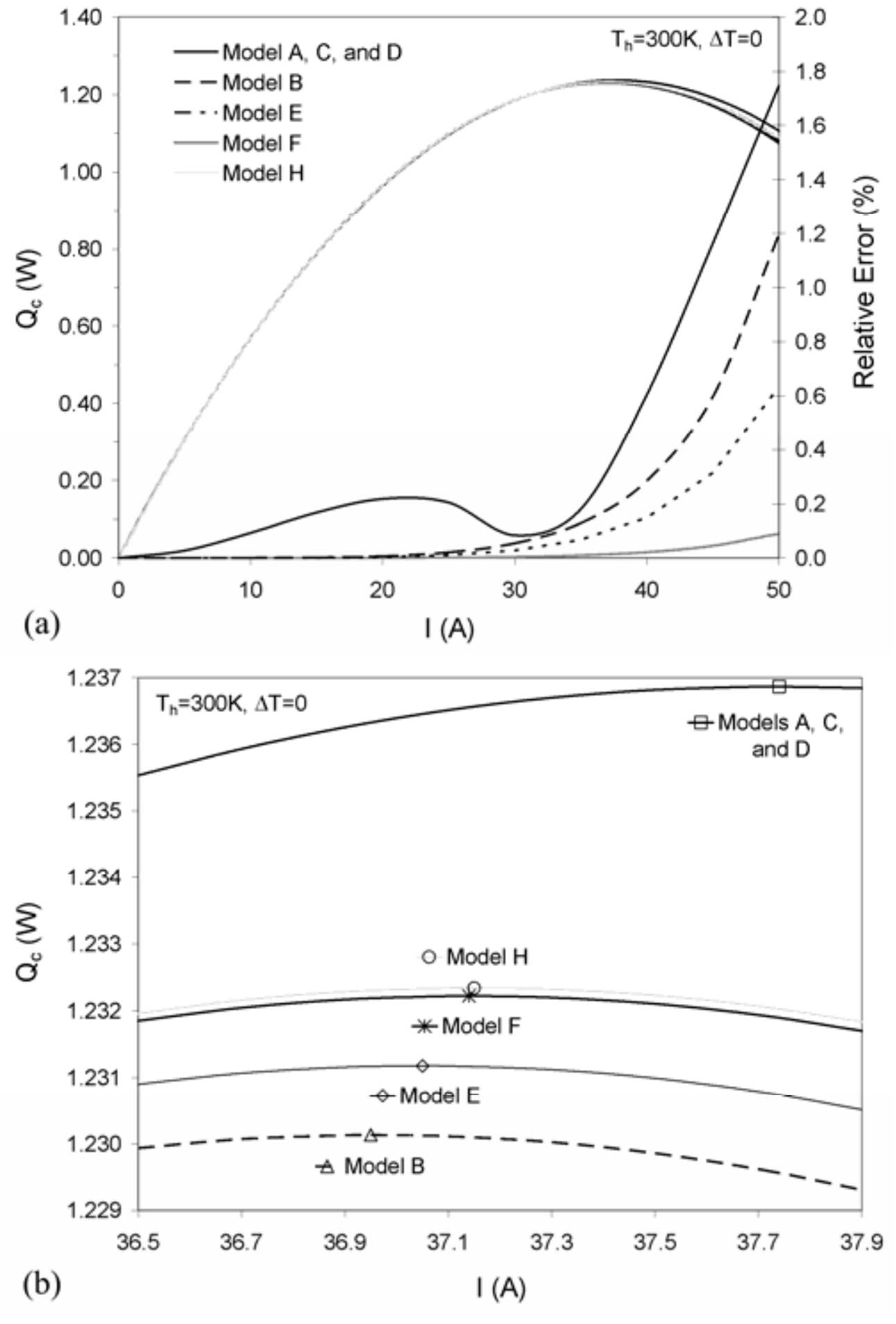

Fig. 7. Prediction comparisons between the different models in Table II for cooling power vs. electrical current at Δ*T*=0ºC, (a) broad electrical current range, including a secondary axis to show the relative error with respect to model H, and (b) expanded data near maximum current.

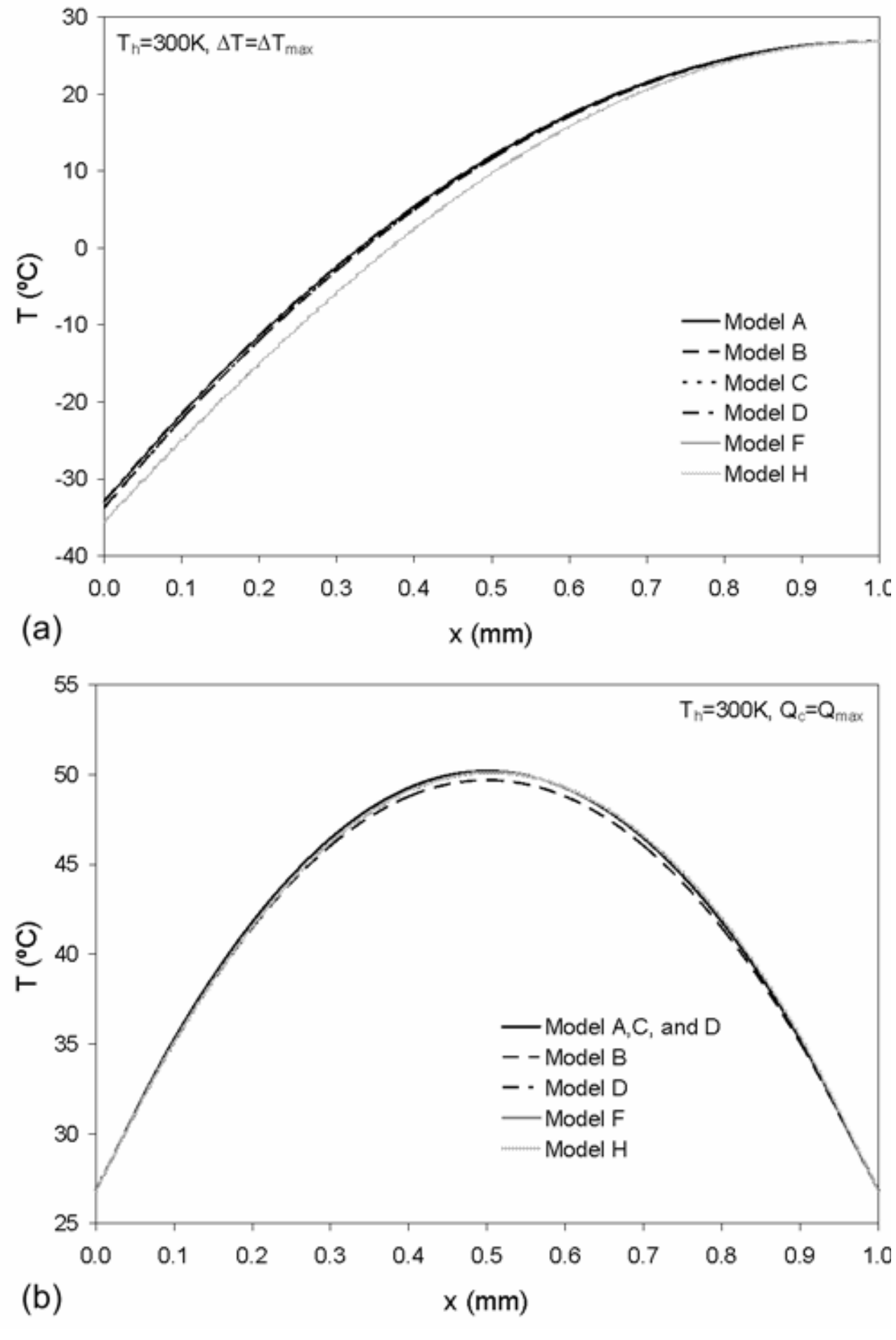

Fig. 8. Comparisons between temperature profile predictions of the different models presented in Table II for the cases (a) when $\Delta T=\Delta T_{max}$., and (b) when $Q_c=Q_{cmax}$.

Finally, spatial profiles of material parameters for the $\Delta T_{max}$ case including the figure-of-merit of a thermoelectric material, $z = s^2 \rho^{-1} \kappa^{-1}$, are shown in Fig 9. These figures clearly illustrate the errors produced when assuming constant material parameters throughout the entire pellet. Furthermore, analyzing Fig. 6(b) and the figure-of-merit profile of Fig. 9(d), it is surprising to see that even though the mean values of *z* across the pellet for numerical simulation (model H) and for distributed parameter (model F) are below the constant values of models B and C, these models actually underestimate the maximum temperature difference predicted by models H and F. In fact, the value $\Delta T_{max}$ obtained with numerical simulation or distributed parameter models is higher than one would expect from even the highest figure-of-merit within the pellet. The explanation is that numerical simulation and distributed parameter models include the effect of Thomson cooling, which is neglected by the lumped parameter models [4,16].

## CONCLUSIONS

Based on one-dimensional steady-state analysis of thermoelectric phenomena and on analogies between thermal and electrical domains, two types of electrical three-port models for thermoelectric devices have been proposed: lumped and distributed parameter. An important advantage of these models is that both electrical and thermal behavior is simulated using a common electronic circuit simulation tool (e.g., SPICE simulation programs), thus allowing analysis of the overall thermoelectric system performance, including control electronics and thermal elements. Lumped parameter models are easily implemented and, if carefully used, can provide accurate results. Distributed parameter models account for temperature dependence of material properties, thus they are more accurate but slightly more complicated to implement. Simulations made for a $(Bi_{0.5}Sb_{0.5})_2Te_3$ pellet show that for the cases of maximum temperature difference, $\Delta T_{max}$, and maximum cooling power, $Q_{cmax}$, a lumped parameter electrical model with material parameters evaluated at mean pellet temperature has, with respect to the highly realistic numeric simulation, a relative error of less than 3.2% for $\Delta T_{max}$ and less than 0.2% for $Q_{cmax}$. By dividing the same pellet into ten segments, the distributed parameter electrical model reduces the relative error to 0.03% for $\Delta T_{max}$ and 0.01% for $Q_{cmax}$.

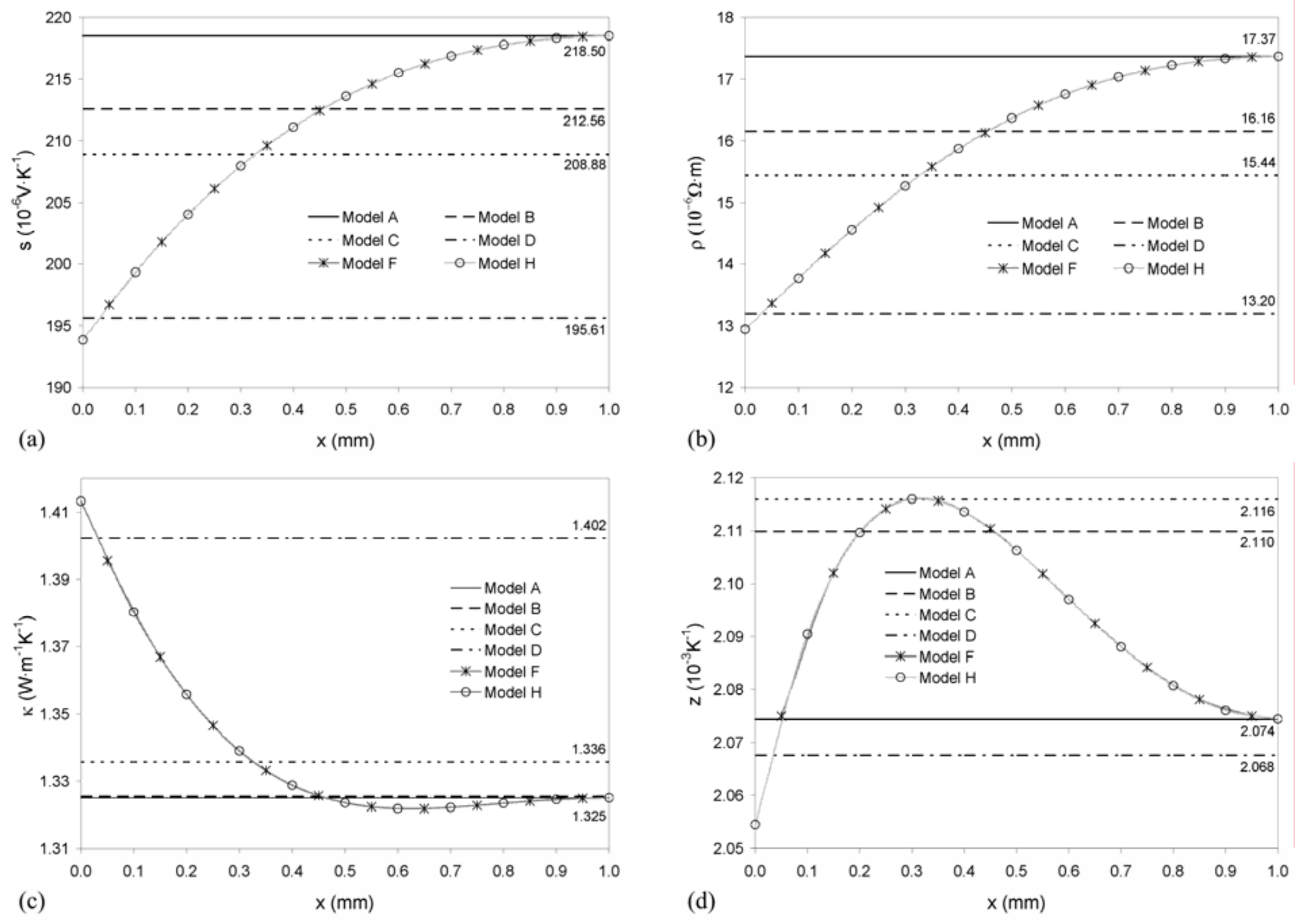

Fig. 9. Comparisons between the different models in Table II of spatial profiles of material parameters $s(x)$, $\rho(x)$, $\kappa(x)$, and $z(x)$ for $\Delta T_{max}$ case (see Fig. 6) according to material data given in [15].